\documentclass[twocolumn,showkeys]{revtex4}
\usepackage[utf8x]{inputenc}
\usepackage[T1]{fontenc}
\usepackage[english]{babel}
\usepackage{graphicx,url,booktabs}

\begin{document}
\title{Tomographic docking suggests the mechanism of auxin receptor TIR1 selectivity}
\author{Veselina V. \surname{Uzunova}}
	\affiliation{School of Life Sciences, University of Warwick, Gibbet Hill Road, Coventry CV4 7AL, UK}
\author{Mussa \surname{Quareshy}}
	\affiliation{School of Life Sciences, University of Warwick, Gibbet Hill Road, Coventry CV4 7AL, UK}
\author{Charo I. \surname{del~Genio}}
	\affiliation{School of Life Sciences, University of Warwick, Gibbet Hill Road, Coventry CV4 7AL, UK}
\author{Richard M. \surname{Napier}}
	\email{Richard.Napier@warwick.ac.uk}
	\affiliation{School of Life Sciences, University of Warwick, Gibbet Hill Road, Coventry CV4 7AL, UK}
\date{\today}

\begin{abstract}
We study the binding of plant hormone IAA on its receptor TIR1 introducing
a novel computational method that we call \emph{tomographic docking} and that
accounts for interactions occurring along the depth of the binding pocket.
Our results suggest that selectivity is related to constraints that potential
ligands encounter on their way from the surface of the protein to their final
position at the pocket bottom. Tomographic docking helps develop specific
hypotheses about ligand binding, distinguishing binders from non-binders,
and suggests that binding is a three-step mechanism, consisting of engagement
with a niche in the back wall of the pocket, interaction with a molecular
filter which allows or precludes further descent of ligands, and binding on
the pocket base. Only molecules that are able to descend the pocket and bind
at its base allow the co-receptor IAA7 to bind on the complex, thus behaving
as active auxins. Analyzing the interactions at different depths, our new
method helps in identifying critical residues that constitute preferred future
study targets and in the quest for safe and effective herbicides. Also, it
has the potential to extend the utility of docking from ligand searches to
the study of processes contributing to selectivity.
\end{abstract}


\maketitle

\section{Introduction}
The molecular recognition of specific small organic compounds
by target proteins is of central importance in biology. Auxins
are a particularly relevant class of small molecule plant hormones
with considerable importance for growth and development. Both
the naturally occurring indole-3-acetic acid~(IAA) and synthetic
auxins bind to the Transport Inhibitor Response~1~(TIR1) family
of receptors. In turn, auxin binding to the receptor allows the
co-receptor IAA7 to bind to the substrate-receptor complex~\cite{Tan07,Cal12}.
Thus, one can say that auxins act as ``molecular glues'' between
partners of the receptor system. The completion of this two-step
mechanism triggers a cascade of events leading to changes in
gene expression~\cite{Nap14}. The macroscopic results of acute
exposure to synthetic auxins are explosive, epinastic growth
followed by plant death. Thus, such compounds have found widespread
application in herbicidal formulations. Further valuable features
of auxin-based herbicides are a long history of safe use and
their selective action against broad-leaved plants, making them
preferred products for the control of weeds in cereal crops and
turf~\cite{Nap03}. However, rational design of novel biologically
active molecules to influence the TIR1 receptor has proved challenging
because the protein recognizes a diverse set of natural and synthetic
ligands~\cite{Cal12}. At the same time, TIR1 is highly selective.
For instance, the native ligand IAA shares many structural features
with its biosynthetic precursor, the indolic amino acid L-tryptophan
(Trp), which, although ubiquitous and present at intracellular
concentrations far in excess of that of IAA, does not elicit
auxin responses~\cite{Lee14}.

A likely reason for this is that inactive compounds
do not bind the receptor in the right location or with the right
orientation, if at all, thus precluding assembly of the co-receptor
complex. Then, knowledge
of the mechanism of interaction for natural ligands is fundamental for the
design of synthetic analogues. Computational methods for molecular docking
have become standard tools in active compound design and discovery~\cite{Kit04,Jor04,Wis08}.
They allow a reduction of the search space, leading to targeted experimental
binding assays, and they are widely used for ligand screening and identification
of binding sites on bioactive targets~\cite{Sta12,Xie15}. Specific examples
of the application of molecular docking are the identification of a genetic
cause of cancer drug resistance~\cite{Kob05}, ligand differentiation between
human oestrogen receptors~\cite{Bol06}, rational drug design for neurodegenerative
diseases~\cite{Cho10}, modelling candidate therapeutic binding to mutated
targets~\cite{Smi12}, and design of highly catalytic artificial metallobioenzymes~\cite{Hys12}.
In all cases the binding site is considered as a single, holistic search
space. In this article, we introduce a new approach to molecular
docking, which we call \emph{tomographic docking}, that we use to
propose an explanation for the discrimination mechanism of small ligands by
the TIR1 receptor.

\begin{figure*}[t]
\centering
\includegraphics[width=0.8\textwidth]{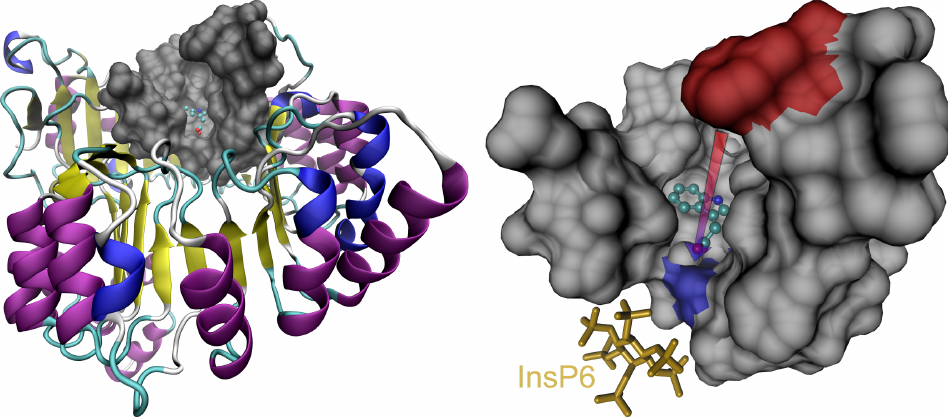}
\caption{The deep binding pocket on TIR1. The binding
site of TIR1 is not a shallow surface indentation but
a deep pocket, shown in SURF representation in the left
panel superimposed on the cartoon of the whole receptor (2P1P).
The residues comprising the binding pocket, isolated
and shown from a closer point in the right panel, are
contributed by seven, non-sequential leucine-rich repeats~\cite{Tan07}.
The two reference residues Phe-351, in red, and Arg-403,
in blue, indicate the mouth and the bottom of the pocket,
respectively. Their distance, indicated by the arrow
in the figure and corresponding to the pocket depth,
is $16.5$Å. IAA is shown in CPK representation, and InsP6
is on the bottom left in bond representation.
All 3d molecular visualizations were produced using VMD~\cite{Hum96,VMD}.}
\label{pocketdef}
\end{figure*}
One frequently overlooked aspect of the molecular recognition process
is the depth of the protein binding site, which can extend significantly
towards the protein core. Several computational tools exist that help
describe and define pockets, tunnels, channels and pores, and some will
identify the most likely high-affinity sites in the recess. Once defined,
these are offered as binding sites for docking. While this approach is
able to find a good candidate for the lowest-energy configuration of
a given receptor-binder pair, it risks neglecting receptor features that
will be encountered by the ligand on the approach to the best site. When
this happens, it contributes to ligand misidentification and false positive
results, both of which are recognized issues with docking experiments~\cite{Xu15}.
For example, AutoDock~Vina~\cite{Tro10,Vina}, which is currently one
of the most popular docking platforms due to its speed, reliability and
output accessibility, finds an apparently viable docking pose for Trp
on TIR1, even though Trp is experimentally proven to be a non-binder.
It is thus reasonable to consider the passage of a ligand into a deep
binding site as a multi-step process composed of many interactions, sequential
in time and space. Consequently, typical docking approaches, which consider
any geometrically valid pose as equally viable, may overlook important
physical and chemical barriers that could preclude some potential binders
from accessing an otherwise ideal site. To take into account the entire
structure of the deep pocket of TIR1, we create a new docking
approach. Rather than considering the whole TIR1 pocket as a single,
whole entity, our method divides it into a number of ``slices'' across
its depth. Each slice is treated individually, so that the results
we obtain in terms of scoring functions and orientations change progressively
with depth. In analogy to the tomographic scans routinely
employed for medical diagnoses, we call our method \emph{tomographic
docking}. The analysis of a whole series of results allows
us to identify physical constraints
that preclude the binding of Trp while allowing that of IAA. Also, we
identify the structural features of the pocket likely to be responsible
for the mechanism of selectivity.

\section{Methods}
\subsection{The target protein}
\begin{figure*}[t]
\centering
\includegraphics[width=0.9\textwidth]{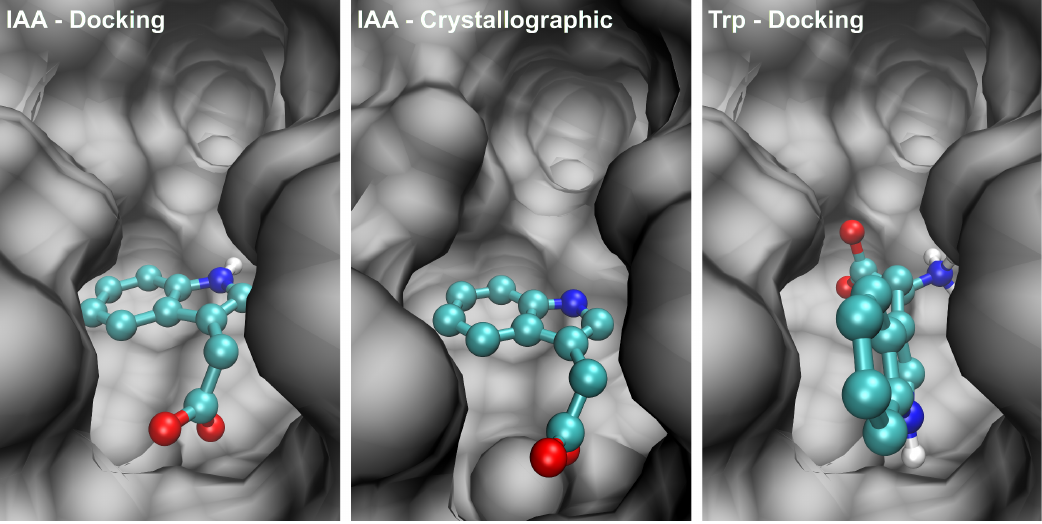}
\caption{Comparison of crystallographic data and ``static'' docking results.
The docked position of IAA at the bottom of the pocket, in the left panel, matches
the crystallographic one in the centre panel. The docked position of Trp at
the bottom of the pocket, in the right panel, does not match the docking and
crystallographic positions of IAA.}
\label{static}
\end{figure*}
The X-ray crystal structure of TIR1 is solved
in three different binding states: unbound/empty (2P1M), in complex
with the natural ligand IAA (2P1P, Fig.\ref{pocketdef}), and assembled
with both IAA and its co-receptor IAA7 (2P1Q). For our study, we
use the unbound structure 2P1M as the closest approximation to what
a free ligand interacts with. Superposition of 2P1M with 2P1P and 2P1Q
suggests no significant conformational change is induced by ligand
binding~\cite{Tan07}.

The crystallography data contain several associated
biomolecules, as well as water. Thus, to prepare the
docking input, we first processed the data using VMD~\cite{Hum96,VMD},
excising water molecules and the co-expressed SCF\textsuperscript{TIR1}
adaptor ASK1. AutoDockTools~\cite{Mor09,ADT} was then
used to produce the final \texttt{pdbqt} input files.
Note that TIR1 harbours inositol hexaphosphate (InsP6)
as a second, probably structural, ligand.
We left this in place because, unlike ASK1, its location
is physically close to the bottom of the pocket. Nonetheless,
our final results show that its position is still too
far from the binding site to generate effective interactions
with the ligand.

\subsection{Ligands}
The investigation focusses on IAA and Trp as TIR1 ligands
because Trp, which is a precursor in the synthesis of IAA~\cite{Woo05},
has no auxin activity and no TIR1-binding activity~\cite{Lee14},
notwithstanding a significant structural similarity with
IAA (Fig.~S1). To validate the accuracy of our approach,
we later extend the analysis to a few other compounds, comprising
both binders and non-binders, with different degrees of structural
similarity to IAA (Fig.~S2). To prepare the ligands for docking,
we first calculated their protonation state at pH $7.3$
in water, since binding to the receptor occurs at physiological
pH in the plant cell nucleus. Then, we computed their equilibrium
geometry using density functional theory with EDF2 functional~\cite{Lin04}
and a 6-31G* split-valence basis. Finally, we generated the
\texttt{pdbqt} files using AutoDockTools~\cite{Mor09,ADT}.

\subsection{Numerical setup}
\begin{figure*}[t]
\centering
\includegraphics[width=0.72\textwidth]{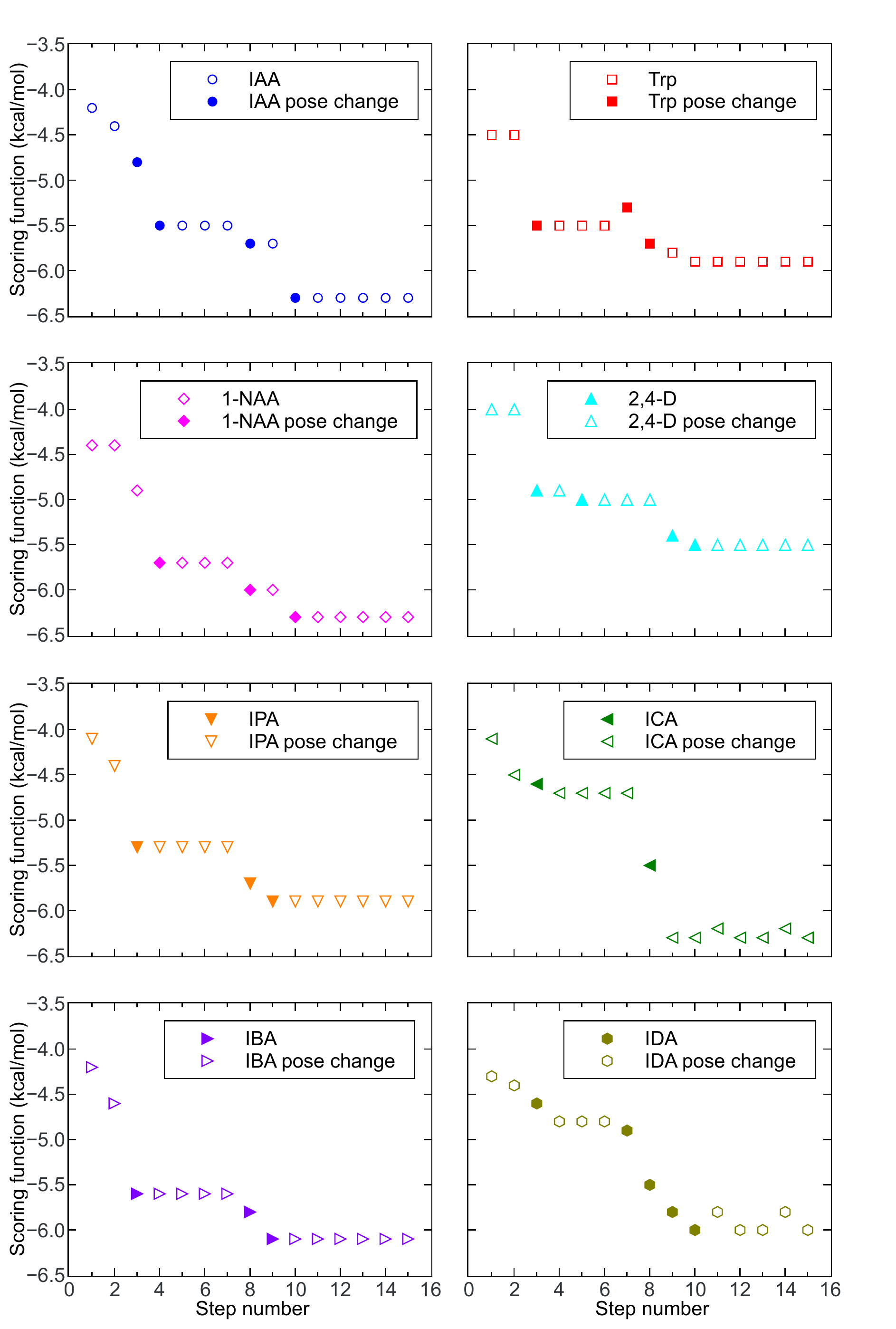}
\caption{Best scores of docked poses along the transect
for all compounds tested, namely IAA (blue circles), Trp
(red squares) 1-NAA (magenta diamonds), 2,4-D (cyan triangles),
IPA (orange triangles), ICA (green triangles), IBA (violet
triangles) and IDA (olive hexagons). Each step represents
the progression of the search space by 1~Å in the direction
of the pocket bottom. Filled symbols indicate steps at
which a significant change in depth or orientation of the
docked pose occurs.}
\label{scorprof}
\end{figure*}
To define the search space to be used in the docking algorithm, we observed
SURF representations of TIR1 in both bound and unbound states, and noted that
the binding site is not a shallow surface feature, but rather a deep binding
pocket (see Fig.~\ref{pocketdef}). In particular, we identified the constituents
of the pocket to be a total of 43 amino acids in seven contiguous sets on the
leucine-rich repeat loops, namely residues 77--84, 344--354, 377--381, 403--410,
436--441, 462--465 and 489--490. The pocket thus defined has a depth of $16.5$~Å,
as measured between Phe-351 at the mouth and Arg-403 at the bottom. To investigate
the engagement process as the ligand moves into it towards the final binding
site at the bottom, we defined an \mbox{18~Å$\times$18~Å$\times$18~Å} cubic search
space that moves from above the pocket mouth to below the bottom in steps of
1~Å. The search space at the first step includes Phe-351 at its bottom,
and its motion is parallel to the principal axis of inertia of the receptor,
whose direction is along the pocket depth.
At the last step, Arg-403 is completely included.
Then, we performed independent numerical docking experiments at each step,
building a sequence of results that provide information on the descent of the
ligands into the deep pocket. For the actual simulations, we created a code
that automates the tomographic scanning process by computing the geometry of
the search space for any specified number of steps, search exhaustiveness,
and set of ligands. The code, which we refer to as TomoDock, uses AutoDock~Vina~\cite{Tro10,Vina}
as docking engine, and produces tunable summaries of the results, as well as
\texttt{pdb} files for further analysis and visualization. Note that with the
choices detailed above, the search space is always larger than Vina's cutoff
threshold for the interactions, which is 8~Å. The standard TomoDock experiment
was repeated 100~times, with search exhaustiveness of~16.

\subsection{Experimental setup}
Experimental evaluation of the numerical results was carried
out using surface plasmon resonance (SPR) as a test of ligand
binding to TIR1, and root growth assays for overall biological
activity. We performed protein purification and set up the SPR
experiments according to the protocols described in~\cite{Lee14}.
TIR1 was expressed in insect cell culture using a recombinant
baculovirus. The construct contained sequences for three affinity
tags, namely 6$\times$His, maltose-binding protein (MBP) and
FLAG. Initial purification using the His tag was followed by
cleavage of His-MBP using TEV protease. After TEV removal and
clean-up using FLAG chromatography, the purified protein was
used for SPR assays by passing it over a streptavidin chip loaded
with biotinylated IAA7 degron peptide. The SPR buffer was Hepes-buffered
saline with 1~mM EDTA, $0.05$\%~P20 and 1mM~TCEP. Compounds
to be tested were premixed with the protein to a final 50~µM
concentration. Binding experiments were run at a flow rate of
20~µl/min using 3~minutes of injection time and $2.5$ minutes
of dissociation time. Data from a control channel (biocytin)
and from a buffer-only run were subtracted from each sensorgram
following the standard double reference subtraction protocol.
To assay root growth inhibition, \mbox{Col-0~WT} seeds were
stratified at 4~°C for 48~hours on plates containing ${}^1/{}_2$
Murashige and Skoog medium, followed by incubation for 6~days
in 12-hour day/night cycles, at a temperature of 20~°C for the
day and 18~°C for the night. Seedlings were then transferred
to fresh plates containing test compound and poured fresh on
the day. After a further 6~days of growth plates were scanned
and the extension of the primary root during treatment was measured
using ImageJ~\cite{Sch15,Image}.

\section{Results and discussion}
\subsection{Conventional docking}
\begin{figure*}[t]
\centering
\includegraphics[width=0.9\textwidth]{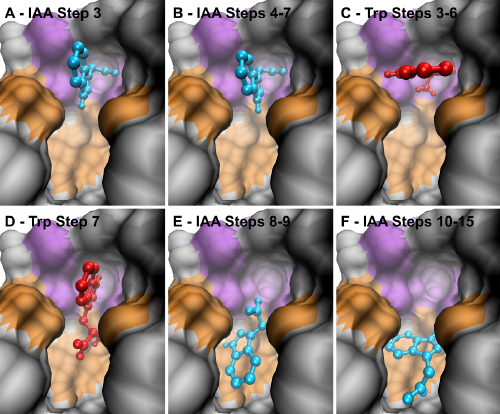}
\caption{Progressive docking poses of IAA (blue) and Trp (red). The residues
that form the engagement niche are highlighted in violet; those that we identify
later as forming a molecular filter are highlighted in orange.
(A) The position of IAA at step~3 features the side chain oriented towards the engagement niche.
(B) The positions of IAA in steps~4 to~7 are superimposable, and show that the ring system is perpendicular
to the bottom of the pocket.
(C) The positions of Trp in steps~3 to~6 are superimposable. Its side-chain
is oriented towards the niche, but the ring system is parallel, rather than
perpendicular, to the pocket base.
(D) At step~7, the side-chain of Trp is no longer in the niche, but points towards the pocket bottom.
(E) At steps~8 and~9 (superimposable), IAA has moved towards the bottom of the pocket.
(F) The positions of IAA at steps~10 to~15 are superimposable.}
\label{poses}
\end{figure*}
We first ran a conventional, ``static'' docking experiment using AutoDock~Vina
using a cubic search space with an 18-Å side encompassing the whole pocket area.
The results show that IAA docks at the pocket bottom and, even from a single run,
the best docked position closely matches that of the ligand bound in the co-crystallised
structure (Fig.~\ref{static}). The indole ring is aligned parallel to the pocket
bottom and is nested in a semi-circle of four non-polar residues, while the carboxylate
anion orients itself towards a group of basic residues with which hydrogen bonds
are made~\cite{Tan07}. Despite the absence of activity for Trp as an auxin and
no measurable affinity for binding, AutoDock~Vina finds an apparently plausible
docked position for it at the bottom of the pocket, although not with
the same orientation as IAA (Fig.~\ref{static}). This clearly shows that one
cannot rely on a direct interpretation of docking results to identify binders,
because a ``cover-all'' search space overlooks key features of the binding process
and a more systemic approach is needed.

\subsection{Tomographic docking}
To study the transient interactions of IAA and Trp with the pocket
as they move down into it, we performed tomographic docking experiments
and analyzed each series of docked poses in detail, building a plausible
binding pathway for both compounds over a transect of 15~Å. The docking
process assigns a lower numerical score to better poses,
representative of lower energy and more favourable binding. Thus, for
each ligand we created a representative series of successive orientations
choosing at each step the pose with the lowest score amongst all repetitions (Fig.~\ref{scorprof}).
Note that the depth at which the ligand is positioned does not necessarily
increase with step number. For instance, as described in greater detail
below, neither the depth, nor the orientation of the docked pose of
IAA changes between step~4 and step~7, indicating that the interactions
relevant over these steps are dominated by the residues included at
step~4, and that no further significant interactions are made until
the ligand approaches residues deeper in the pocket than those at
step~7. Later, we use these considerations to identify which residues
are most likely to be responsible for the selection mechanism.

At step~1 both compounds are well out of the pocket, and at step~2
they are at the very edge of the pocket mouth. As the steps continue,
the progressive inclusion of residues causes the docked position of
IAA to undergo significant changes. At step~3, IAA has oriented itself
with the carboxylic acid group in a niche at the back wall of the
pocket~(Fig.~\ref{poses}A). Then, for the next four steps, its scoring
function, position and orientation remain constant, with the alignment
of the indole perpendicular to the base of the pocket (Fig.~\ref{poses}B).
Note that between step~3 and steps~4--7, IAA undergoes a small but
significant rotation, which optimizes the perpendicularity of the
indole-ring system with respect to the bottom of the pocket.

Considering Trp, its side-chain also becomes oriented
towards the niche at step~3, and its docked depth and orientation
do not change through step~6. However, unlike IAA, the orientation
of the indole is parallel, not perpendicular, to the base
of the pocket. This is probably due to the longer side chain
and the extra rotational degree of freedom, as well as to the proximity
of the aromatic rings to residues distal to the niche (Fig.~\ref{poses}C).
The next step for Trp, step~7, presents a somersault for
the pose, with its polar side groups now pointing towards
the pocket bottom (Fig.~\ref{poses}D).

From the pose in step~7, with the tail in the niche,
IAA can proceed downwards, into the position observed
at step~8, via a pivoting motion of the indole from
the engagement niche (cf.\ Fig.~\ref{poses}B and Fig.~\ref{poses}E).
Poses~8 and~9 for IAA are identical, with the side-chain
continuing to point towards the niche, but not in it.
Then, there appears to be a final transition as residues
at the base of the pocket come into play, with poses~10
to~15 showing that IAA has flipped over from poses~8 and~9
(Fig.~\ref{poses}F), to a position that corresponds
to that found in the crystal structure~\cite{Tan07}.

\subsection{Binding mechanism}
When the docking
algorithm explores positions that include
the pocket bottom, Trp is docked at
the binding site. This indicates that,
in principle, the final docking position
is allowed. However, a detailed examination
of the tomographic docking results
suggests the presence of a barrier impeding
the descent of the non-binder into
the binding pocket, explaining why,
in nature, Trp never reaches its bottom.

\begin{figure}[t]
\centering
\includegraphics[width=0.45\textwidth]{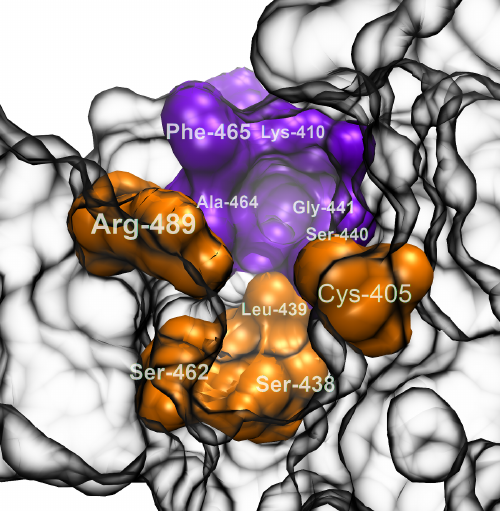}
\caption{Molecular filter responsible for TIR1 selectivity.
The residues shown in orange are responsible
for the filtering mechanism. The engagement niche is shown in purple. The remainder
of the pocket is represented with partial transparency for
clarity.}
\label{filter}
\end{figure}

In the initial part of the pocket the tomographic docking identifies a region
that we call the engagement niche formed by residues Lys-410,
Ser-440, Gly-441, Ala-464 and Phe-465 (in violet in Fig.~\ref{filter}). This is the structure
into which the binder orients its polar side-chain (Fig.~\ref{poses}B).
We deem it one of the features with which potential binders
need to interact in order to achieve an orientation that
allows a subsequent transition to the binding position.
TomoDock results suggest that
the particular orientation with the ring system perpendicular
to the pocket base is likely to be an essential step in
the selection process. For ligands to penetrate deeper,
the aromatic rings must slice down while the polar tail,
anchored in the engagement niche, acts as a pivot point
(cf.\ Fig.~\ref{poses}B and Fig.~\ref{poses}E). After
this motion, the rings of IAA are positioned at the bottom
of the pocket, in the vicinity of a semi-circle of non-polar
residues. The ligand then undergoes a slight rotation,
allowing the polar carboxylic acid group to flip and engage
with the polar residues at the pocket base.

\begin{figure*}[t]
\centering
\includegraphics[width=0.9\textwidth]{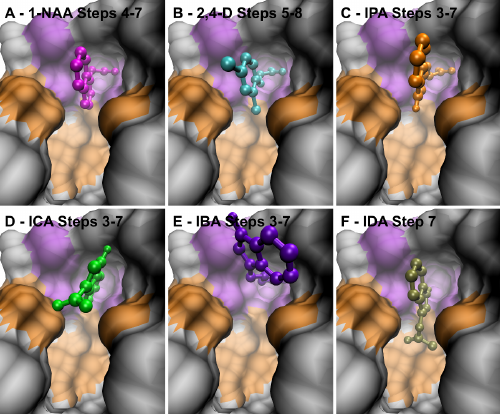}
\caption{Docking poses of 1-NAA (magenta), 2,4-D (cyan), IPA (orange),
ICA (green), IBA (violet) and IDA (olive).
(A) The positions of 1-NAA at steps~4 to~7 (superimposable) show the side-chain of the ligand engaged
with the niche, and its ring system perpendicular to the pocket bottom.
(B) In steps~5 to~8 (superimposable) the orientation of 2,4-D is the same as that of IAA and 1-NAA.
(C) At steps~3 to~7, also IPA engages the niche with the ring system perpendicular to the pocket base. The positions at these steps are superimposable.
(D) In steps~3 to~7 (superimposable), the tail of ICA never engages the niche.
(E) The tail of IBA finds the niche in steps~3 to~7 (superimposable), but the ring system is misoriented.
(F) At step 7, the side-chain of IDA is oriented towards the pocket bottom, in a pose reminiscent of that of Trp at the same depth.}
\label{validposes}
\end{figure*}
To understand what blocks Trp from moving the same way as IAA, consider
the results from steps~3 to~7. At step~3, Trp is docked with its polar tail in
the engagement niche. However, we do not observe the perpendicular orientation
of the aromatic system that we see in IAA (Fig.~\ref{poses}C). Note that its orientation and
docking depth do not change through step~6 (Fig.~\ref{scorprof}). Then, at step~7, Trp assumes
a new pose with the side-chain completely out of the niche, and pointing
towards the pocket bottom (Fig.~\ref{poses}D). Such a geometry prevents
the non-binder from moving further into the binding position via the same
rotation that IAA performs, due to inappropriate orientation of the indole rings
and of the side chain.

Note that the non-binder is allowed to dock further down the pocket
from step~8 onwards (Fig.~\ref{scorprof}), because docking considers any position that
is geometrically accessible, disregarding the motion a ligand would
have to undertake in order to reach it.
Also, to be active auxins, substrates
need to bind in the correct orientation at the bottom of the
pocket. A compound that can only interact and bind at the
mouth of the pocket cannot have auxinic activity. Thus, for brevity
we refer to compounds that are able to achieve an appropriate
binding position simply as ``binders''. Conversely, we refer to
the compounds, like Trp, that cannot reach the pocket bottom as ``non-binders''.

\subsection{Validation and experimental verification}
\begin{figure*}[t]
\centering
\includegraphics[width=0.34\textwidth]{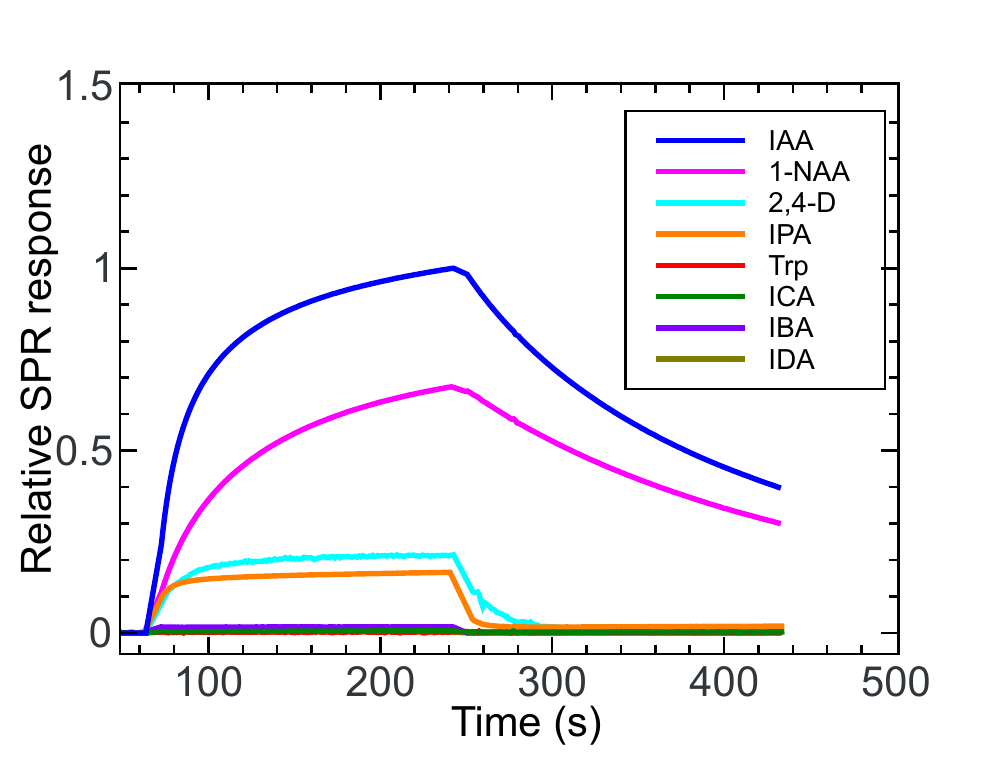}
\includegraphics[width=0.62\textwidth]{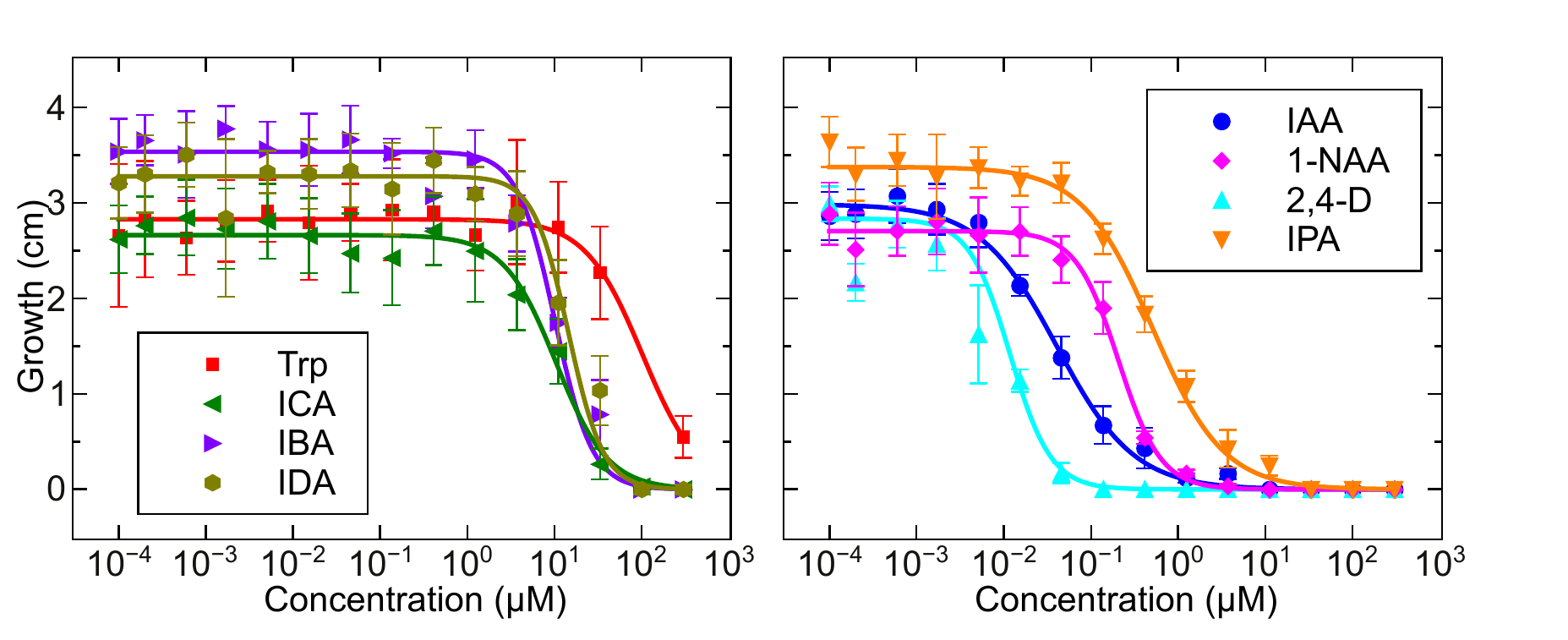}
\caption{Surface plasmon resonance and root growth assay results confirm the numerical predictions.
The results from SPR experiments (left panel) show that Trp (red line), ICA (green line), IBA
(violet line) and IDA (olive line) do not bind TIR1 at all. Conversely, 1-NAA (magenta line),
2,4-D (cyan line) and IPA (orange line) all bind, although with differing activities from IAA (blue
line). All compounds were tested at 50~µM. Root growth inhibition measurements (centre and right panels)
substantially confirm these
results, revealing that Trp (red squares), ICA (green triangles), IBA (violet triangles)
and IDA (olive hexagons) do not inhibit root growth up to extreme concentrations. In the right-hand
panel, 1-NAA (magenta diamonds), 2,4-D (cyan triangles) and IPA (orange triangles) are all active
auxins. Derived IC50 values are given in Table~\ref{ic50}.}
\label{exper}
\end{figure*}
The striking difference between
IAA and Trp revealed by tomographic docking
indicates that an important role is
played by the residues that become available
at steps~3, 4, 7, 8 and~10 (filled symbols in
Fig.~\ref{scorprof}). In particular, the TomoDock
results suggest that they act as a molecular filter, promoting
the correct orientation of IAA, and opposing
it for Trp. To identify these residues, we built
a table of the atoms newly included for interaction
at each step, along with the residues they belong
to (Table~S1). Then, we considered the new entries
at the steps indicated above, taking
into account the number of atoms that interact,
as well as their properties.

As a first example,
consider Ser-438 (see Fig.~\ref{filter}). This residue enters the search
space at step~10, which is the first step at which
IAA assumes its final binding position. Ser-438 is
physically located at the bottom of the pocket
and, upon close inspection, the atoms that
get included at step~10 are seen to form a highly polar
group. Thus, we include it in the molecular
filter, and consider it responsible for the
correct orientation of IAA at the pocket bottom.

As a second example, consider Ser-440. This
residue is structurally part of the engagement
niche, and it enters the search space at step~6,
with~3 atoms. However, neither IAA nor Trp
changes its position at all over this step (see
Fig.~\ref{poses}B and~C). At step~7, where
4~more atoms of Ser-440 are considered,
Trp exits the niche (Fig.~\ref{poses}D).
One could thus consider Ser-440 partly responsible
for this; however, at step~7 IAA maintains the
same position as it has at step~6. Given the
structural similarity of the two molecules,
we believe this indicates that Ser-440 does
not contribute actively to ligand filtering,
particularly considering
that a much better candidate for the observed
effect on Trp exists, namely Leu-439.

The third example we discuss
is Gly-441. This is a noteworthy residue,
as it is not only structurally part of
the engagement niche, but also because
its mutation to aspartate yields the known
tir1-2 mutant~\cite{Rue98}. This residue
gives its first big conribution to the
search space at step~4. But by this step
both IAA and Trp have already assumed
positions that do not change for a
few more steps. Thus, we do not consider
this residue as an active player
in the molecular
filter.
Substitution of the large polar side group
of Asp for the small non-polar Gly could
interfere with binding in many ways to explain
the tir1-2 phenotype.

Performing this analysis on all viable
residues shows that the filter is formed by Cys-405,
Ser-438, Leu-439, Ser-462 and Arg-489 (in orange in Fig.~\ref{filter}).
Of these, Arg-489 seems to be the residue
that most significantly affects the orientation
of the compounds with respect to the engagement
niche, of which, however, it is not a part.
Leu-439 and Ser-462
appear to block the descent of Trp
and promote that of IAA, as they progressively
enter the search space in steps~7--9. Finally,
Cys-405 and Ser-438 are likely to be instrumental
for IAA to assume the final binding position,
since they start contributing significantly
at steps~8 and~10, respectively. Note that all
the residues in physical
proximity of the ligand at the pocket bottom are likely to
be involved in stabilizing docked auxins,
but they are not necessarily part of
the filtering mechanism.

To further exemplify our method, we apply it to six other potential
binders (Fig.~S2), namely indole-3-carboxylic acid (ICA), indole-3-propionic
acid (IPA), indole-3-butyric acid (IBA), indol-3-yl acetate (IDA), 1-naphthaleneacetic
acid (1-NAA) and 2,4-dichlorophenoxyacetic acid (2,4-D). The aliphatic
side-chains of ICA, IPA and IBA differ in length from that of IAA,
being one atom shorter, one atom longer and two atoms longer,
respectively. Esterifying indole-3-ol with acetic acid gives
IDA, while changing the indole system to a naphtalene double ring yields
1-NAA. Finally, we include 2,4-D as it is one of the most widely used
herbicides in the world.
For all compounds, tomographic docking predicts a plateaux between
steps~3 and~7 (Fig.~\ref{scorprof}). At these steps, all compounds are docked at the right
depth to permit interaction with the engagement niche.
Results for 1-NAA, 2,4-D and IPA show that they do engage with
the niche, with the aromatic ring system in the same position as that
of IAA, perpendicular to the base of the pocket (Fig.~\ref{validposes}A--C).
In contrast, the results for ICA, IBA
and IDA predict that they are not able to adopt the correct engagement pose
to allow the subsequent pivot. Whilst they may occupy space in and interact
with the outer chamber, TomoDock suggests that they fail to orient
appropriately for transit further down the pocket. In the case
of ICA and IDA, there is no
interaction with the engagement
niche, while IBA orients its ring systems transversal, rather
than perpendicular, to the pocket bottom,
similar to Trp (Fig.~\ref{validposes}D--F).
These poses are consistent with 1-NAA, 2,4-D and IPA being active ligands for
TIR1, albeit with different affinities, and with ICA, IBA and IDA being not.

Experimental confirmation of binding from SPR and root growth assays,
(Fig.~\ref{exper} and Tab.~\ref{ic50}), support the numerical predictions.
In particular, SPR experiments indicate that Trp, ICA,
IBA and IDA have no measurable binding activity at 50~µM,
a concentration far in excess of the IC50 value of all active auxins (Tab.~\ref{ic50}).
IPA and 2,4-D, instead, bind weakly compared to IAA and 1-NAA,
with noticeably more rapid dissociation rates (see Fig.~\ref{exper}
and related results in~\cite{Cal12}). Like IAA,
1-NAA is a strong ligand.
Notice that, as mentioned above, substrate binding to the receptor
is only the first step in the auxinic interaction,
and the binding of the co-receptor IAA7 (in our case) to the substrate-receptor
complex can only happen if the substrate is bound in
the bottom of the receptor pocket, and in the correct orientation.
Thus, SPR experiments offer a good method
to validate the computational results: if a
substrate binds incorrectly, it will impair or entirely preclude
the binding of the receptor to the IAA7 co-receptor and
produce no SPR signal.
The relative effectiveness of each compound in root growth assays
compares favourably with the SPR measurements.
Trp, ICA, IBA and IDA inhibit root growth only at very high
concentrations, where phytotoxicity sets in. IPA is seen to be a weak auxin.
The root growth assays suggest that 2,4-D is the most active auxin,
more active than the SPR data suggest. However,
root growth assay activities depend
on tissue
and cellular transport, as well as on receptor binding
of the compounds in question, and so IC50 values do not
correspond exactly to in vitro binding values.
Nonetheless, the assays are still a useful
verification method, as one cannot observe root-growth
inhibition for compounds that do not correctly bind to
the receptor.

\section{Conclusions}
We have studied the binding process of the plant
growth hormone IAA onto its main receptor TIR1,
to investigate its selectivity
mechanism. For our study, we developed a novel tomographic docking
approach suitable for investigating deep binding pockets in a series
of pseudo-time  steps. The method gradually changes the search space
of a docking algorithm to allow one to consider sequential interactions
of each potential ligand with pocket residues at increasing depths. This mimics
what happens in nature when a small molecule descends into a binding
cavity.
In the present study we have considered
the receptor structure as rigid, as is the case for most docking
experiments. However
tomographic docking can be adapted to allow for the flexibility
of some side-chains of the receptor,
and this will be the subject
of future work.
The tomographic method shows a plateau of scoring function
values part-way down the pocket, indicating a region over which
transient interactions are made en route to the docking site at
the base.

Detailed study of the docking
poses obtained for the natural ligand IAA and the related,
but non-auxin Trp over this region points towards two features
in the pocket
responsible for selectivity. The first, an engagement niche in
the back of the pocket, allows potential ligands to orient before
subsequent motion towards the binding site. The second,
a molecular filter, promotes the correct pose of the aromatic ring
system for binders, necessary to access the pocket bottom.
Tryptophan and a set of non-binders assume
sub-optimal orientations, and are prohibited from
onwards motion.

\begin{table}[b]
\begin{tabular}{@{}lr@{.}l@{±}r@{.}l@{}}\toprule
 Compound & \multicolumn{4}{c}{IC50 (µM)}\\ \midrule
 2,4-D & 0  & 0118                      & 0 & 0009\\
 IAA   & 0  & 041                       & 0 & 007\\
 1-NAA & 0  & 206                       & 0 & 015\\
 IPA   & 0  & 51                        & 0 & 06\\
 IBA   & 10 & 1                         & 1 & 2\\
 ICA   & 10 & 4                         & 1 & 2\\
 IDA   & \multicolumn{1}{r@{$\;$}}{15}  &   & \multicolumn{1}{r@{$\;$}}{2}  & \\
 Trp   & \multicolumn{1}{r@{$\;$}}{102} &   & \multicolumn{1}{r@{$\;$}}{13} & \\ \bottomrule
\end{tabular}
\caption{IC50 for primary root growth inhibition derived from root growth assays.}\label{ic50}
\end{table}
The identification of the residues that form the engagement
niche and the molecular filter makes them a fundamental study
subject for the rational design of novel auxin-based herbicides.
They are critical for selectivity, and constitute preferred
mutation targets for further experiments. One
such mutant, tir1-2 (G411-Asp),
is already known, and
experiments have shown that the mutation has a small
but measurable effect on the resistance of the plant
to the auxin-transport inhibitor 2-carboxyphenyl-3-phenylpropane-1,2-dione
(CPD)~\cite{Rue98}, although this non-conservative
substitution may not help elucidate the role of
residue~441 further.

Experimental results from SPR
and root growth assays performed on a set of active and inactive compounds
are consistent with TomoDock results, confirming the validity
of our method. 
The application of tomographic docking
need not be limited to the analysis of auxin binding
to TIR1. In fact, it can be used to examine also other members
of the TIR/AFB auxin receptor family. Identifying similarities
and differences between the interaction mechanisms in
different receptors can play a key role in designing
receptor-specific compounds, which are very useful in
controlling herbicide resistance. In addition,
the tomographic
docking principle is general, and can be applied to any deep binding
site. Thus,
proteins such as those involved in the transport of small
molecules, as well as enzymes and channel proteins,
are all natural targets for tomographic
docking investigation.

\section*{Data accessibility}
The protein structures used in this study are available from the Protein
Data Bank web site at the URL \url{http://www.rcsb.org/pdb/home/home.do}
The TomoDock code is available on the web pages of the authors.

\section*{Competing interests}
We have no competing interests.

\section*{Authors' contributions}
VVU and CIDG performed the numerical simulations.
MQ designed and performed all the SPR and bioassay
experiments. CIDG implemented the TomoDock code.
All authors developed the principle of tomographic
docking, from an original idea of VVU. All authors
analyzed data and results, and wrote the manuscript.

\section*{Funding}
VVU and RMN acknowledge support from BBSRC via grant n.~BB/L009366.
MQ acknowledges support from BBSRC via a studentship awarded through
the Midlands Integrative Biosciences Training Partnership (MIBTP).


\begin{thebibliography}{99}
\bibitem{Tan07} Tan X, Calderón-Villalobos LIA, Sharon M, Zheng CX, Robinson CV, Estelle M, Zheng N. 2007 Mechanism of auxin perception by the TIR1 ubiquitin ligase. \textit{Nature} \textbf{446} 640--645. (doi:10.1038/nature05731)
\bibitem{Cal12} Calderón Villalobos LIA et al. 2012 A combinatorial TIR1/AFB-Aux/IAA co-receptor system for differential sensing of auxin. \textit{Nat. Chem. Biol.} \textbf{8}, 477--485. (doi:10.1038/nchembio.926)
\bibitem{Nap14} Napier RM. 2014 Auxin Receptors and Perception. In \textit{Auxin and Its Role in Plant Development.} E. Zažímalová, J. Petrášek and E. Benková, editors. Springer-Verlag, Vienna, 101--116. (doi:10.1007/978-3-7091-1526-8)
\bibitem{Nap03} Napier RM. 2003 Regulators of growth: Auxins. In \textit{Encyclopedia of Applied Plant Sciences.} B. Thomas, D. J. Murphy, B. G. Murray, editors. Academic Press, Waltham, MA, 985--995.
\bibitem{Lee14} Lee S, Sundaram S, Armitage L, Evans JP, Hawkes T, Kepinski S, Ferro N, Napier RM. 2014 Defining Binding Efficiency and Specificity of Auxins for SCF\textsuperscript{TIR1/AFB}-Aux/IAA Co-receptor Complex Formation. \textit{ACS Chem. Biol.} \textbf{9} 673--682. (doi:10.1021/cb400618m)
\bibitem{Kit04} Kitchen DB, Decornez H, Furr JR, Bajorath J. 2004 Docking and scoring in virtual screening for drug discovery: Methods and applications. \textit{Nat. Rev. Drug Discov.} \textbf{3} 935--949. (doi:10.1038/nrd1549)
\bibitem{Jor04} Jorgensen WL 2004 The many roles of computation in drug discovery. \textit{Science} \textbf{303} 1813--1818. (doi:10.1126/science.1096361)
\bibitem{Wis08} Wishart DS, Knox C, Guo AC, Cheng D, Shrivastava S, Tzur D, Gautam B, Hassanali M. 2008 DrugBank: a knowledgebase for drugs, drug actions and drug targets. \textit{Nucleic Acids Res.} \textbf{36} D901--D906. (doi:10.1093/nar/gkm958)
\bibitem{Sta12} Stark JL, Powers R. 2012 Application of NMR and Molecular Docking in Structure-Based Drug Discovery. In \textit{NMR of Proteins and Small Biomolecules.} G. Zhu, editor. Springer-Verlag, Berlin, 1--34. (doi:10.1007/128\_2011\_213)
\bibitem{Xie15} Xie Z-R, Hwang M-J. 2015 Methods for Predicting Protein-Ligand Binding Sites. In \textit{Molecular Modeling of Proteins.} A. Kukol, editor. Springer Science, New York, NY, 383--398. (doi:10.1007/978-1-4939-1465-4\_17)
\bibitem{Kob05} Kobayashi S, Boggon TJ, Dayaram T, Janne PA, Kocher O, Meyerson M, Johnson BE, Eck MJ, Tenen DG, Halmos B. 2005 EGFR mutation and resistance of non-small-cell lung cancer to gefitinib. \textit{New Engl. J. Med.} \textbf{352} 786--792. (doi:10.1056/NEJMoa044238)
\bibitem{Bol06} Bologa CG et al. 2006 Virtual and biomolecular screening converge on a selective agonist for GPR30. \textit{Nat. Chem. Biol.} \textbf{2} 207--212. (doi:10.1038/nchembio775)
\bibitem{Cho10} Choi JS, Braymer JJ, Nanga RPR, Ramamoorthy A, Lim MH. 2010 Design of small molecules that target metal-A$\beta$ species and regulate metal-induced A$\beta$ aggregation and neurotoxicity. \textit{Proc. Natl. Acad. Sci. USA} \textbf{107} 21990--21995. (doi:0.1073/pnas.1006091107)
\bibitem{Smi12} Smith CC et al. 2012 Validation of ITD mutations in FLT3 as a therapeutic target in human acute myeloid leukaemia. \textit{Nature} \textbf{485} 260--U153. (doi:10.1038/nature11016)
\bibitem{Hys12} Hyster TK, Knorr L, Ward TR, Rovis T. 2012 Biotinylated Rh(III) Complexes in Engineered Streptavidin for Accelerated Asymmetric C-H Activation. \textit{Science} \textbf{338} 500--503. (doi:10.1126/science.1226132)
\bibitem{Xu15}  Xu W, Lucke AJ, Fairie DJ. 2015 Comparing sixteen scoring functions for predicting biological activities of ligands for protein targets. \textit{J. Mol. Graphics and Model.} \textbf{57} 76-88. (doi:10.1016/j.jmgm.2015.01.009)
\bibitem{Tro10} Trott O, Olson AJ. 2010 AutoDock Vina: improving the speed and accuracy of docking with a new scoring function, efficient optimization and multithreading. \textit{J. Comput. Chem.} \textbf{31} 455–461. (doi:10.1002/jcc.21334)
\bibitem{Vina}  AutoDock Vina \url{http://vina.scripps.edu/}
\bibitem{Sch15} Schindelin J, Rueden CT, Hiner MC, Eliceiri KW. 2015 The ImageJ ecosystem: An open platform for biomedical image analysis. \textit{Mol. Reprod. Dev.} \textbf{82} 518--529.
\bibitem{Image} ImageJ \url{http://imagej.net/}
\bibitem{Hum96} Humphrey W, Dalke A, Schulten K. 1996 VMD - Visual Molecular Dynamics. \textit{J. Molec. Graphics.} \textbf{14} 33--38. (doi:10.1016/0263-7855(96)00018-5)
\bibitem{VMD}   VMD - Visual Molecular Dynamics \url{http://www.ks.uiuc.edu/Research/vmd/}
\bibitem{Mor09} Morris GM, Huey R, Lindstrom E, Sanner MF, Belew RK, Goodsell DS, Olson AJ. 2009 Autodock4 and AutoDockTools4: automated docking with selective receptor flexibility. \textit{J. Comput. Chem.} \textbf{16} 2785--2791. (doi:10.1002/jcc.21256)
\bibitem{ADT}   AutoDock \url{http://autodock.scripps.edu/resources/adt}
\bibitem{Woo05} Woodward AW, Bartel B. 2005 Auxin: Regulation, action, and interaction. \textit{Ann. Bot. - London} \textbf{95} 707--735. (doi:10.1093/aob/mci083)
\bibitem{Lin04} Lin CY, George MW, Gill PMW. 2004 EDF2: A Density Functional for Predicting Molecular Vibrational Frequencies. \textit{Aust. J. Chem.} \textbf{57} 365--370. (doi:10.1071/CH03263)
\bibitem{Rue98} Ruegger M, Dewey E, Gray WM, Hobbie L, Turner J, Estelle M. 1998 The TIR1 protein of Arabidopsis functions in auxin response and is related to human SKP2 and yeast Grr1p. \textit{Gene. Dev.} \textbf{12} 198--207. (doi:10.1101/gad.12.2.198)
\end{thebibliography}
\end{document}